\documentclass[twocolumn]{svjour3}         
\smartqed  
\usepackage{graphicx}
\usepackage{mathptmx}      
\usepackage{latexsym}
\usepackage{amsmath}
\usepackage{amssymb}
\usepackage{natbib}
\usepackage[usenames]{color}
\usepackage{ulem}

\newcommand{\comment}[1]{}

\journalname{Theor Appl Climatol}
\begin{document}

\title{Fractal characterization of rain-gauge networks and precipitations: 
       an application in Central Italy
}


\author{Valerio Capecchi        \and
        Alfonso Crisci          \and
        Samantha Melani         \and
        Marco Morabito          \and
        Paolo Politi            
}


\institute{V. Capecchi \at
              Institute of Biometeorology \& LaMMA\\
              Via Madonna del Piano 10,
              Sesto Fiorentino, Florence, Italy
              \\\email{v.capecchi@ibimet.cnr.it}
           \and
           S. Melani \at
              Institute of Biometeorology, Via Madonna del Piano 10,
              Sesto Fiorentino, Florence, Italy
           \and
           A. Crisci \at
              Institute of Biometeorology, Via Caproni 6, Florence, Italy
           \and
           M. Morabito \at
              Interdepartmental Center of Bioclimatology, University of Florence, 
              Piazzale delle Cascine 18, Florence, Italy
           \and
           P. Politi \at
              Istituto dei Sistemi Complessi, Consiglio Nazionale delle Ricerche,
              Via Madonna del Piano 10, Sesto Fiorentino, Italy
}

\date{Received: date / Accepted: date}

\maketitle

\begin{abstract}
The measuring stations of a geophysical network are often
spatially distributed in an inhomogeneous manner. 
The areal inhomogeneity  
can be well characterized by the fractal dimension
$D_H$ of the network, which is usually smaller than the euclidean dimension 
of the surface, this latter equal to 2.
The resulting dimensional deficit, $(2-D_H)$, is a measure of
precipitating events which cannot be detected by the network.
The aim of the present study is to estimate
the fractal dimension of a rain-gauge network 
in Tuscany (Central Italy) and to relate its dimension
to the dimensions of daily rainfall events detected by a mixed satellite/radar
methodology.
We find that $D_H\simeq 1.85$, while typical summer precipitations are
characterized by a dimension much greater than the dimensional deficit 0.15.
\end{abstract}
\section{Introduction} \label{intro}
The distribution of a geophysical network is a 
multi-stage decision process which mainly relies on
economic and demographic interests and on access problems
in remote areas. Although an ideal network of stations
should be spatially homogeneous 
 and sufficiently
dense to discriminate the minimum wave-length
of the investigated geophysical phenomena, 
the irregularity and sparsity of observation points 
imply interpolation errors when reporting data on a regular grid.
The areal clustering of point-sets can be
measured by statistical indices as pointed out by \cite{ouchi1986} or, 
when the inter-station distances are scale-invariant, it can be well
characterized thanks to a fractal analysis (\citeauthor{mandelbrot1982} 
\citeyear{mandelbrot1982}).
\newline If the point-set is self-similar, i.e. any small part
of it is the magnified version of the whole set, the set is
a fractal and it can be characterized by its fractal dimension $D_H$, 
which is a real number with $D_H < D_{E}$, where 
$D_{E}$ is the standard euclidean dimension of the embedding space
(in our case, $D_E=2$).
In literature several works
(\citeauthor{korvin1990} \citeyear{korvin1990};
\citeauthor{lovejoy1986} \citeyear{lovejoy1986}; 
\citeauthor{mazzarella2000} \citeyear{mazzarella2000};
\citeauthor{olsson1996} \citeyear{olsson1996}; 
\citeauthor{tessier1994} \citeyear{tessier1994})
deal with the fractal characterization
of a single-point observation network and sometimes this analysis is used as a method to 
drive an optimal enlargement of the network (\citeauthor{mazzarella2000} 
\citeyear{mazzarella2000}).
\cite{lovejoy1986} state that any sufficiently sparsely distributed 
phenomena having a fractal dimension smaller than the dimensional deficit 
$\delta = 2 -D_H$
of the
observing network cannot be detected by the network itself.
Since the sparse precipitating phenomena are the most intense and potentially severe, they are of
prominent interest, particularly when the network of measuring
stations are constantly used for civil protection purposes.
The aim of this study was to compute the fractal dimension $D_H$ of the
rain-gauge network belonging to the Centro Funzionale Regionale in Tuscany (Central Italy)
and to compare $D_H$ with
the fractal dimension $d$ of daily rainfall events occurring in the same area,
using an independent network for rainfall, for the month of July 2010.
We find that $D_H\simeq 1.85$, therefore giving a dimensional deficit $\delta\simeq 0.15$.
On the other hand, all rain patterns give a fractal dimension $d>0.6$, well above $\delta$.
A rough extrapolation of data for $d$ as a function of 24-hours rain thresholds suggests
that our rain-gauge networks might fail to record precipitation events whose
intensity is about 75 mm/day or more.
\par The paper is organized as follows. In section \ref{sec:datamethods} 
we detail the data used in this study and the methodology adopted to evaluate 
$D_H$.
In section \ref{sec:results} the computation of the fractal dimension of the rain-gauge network 
is presented and compared to those obtained for rainy days.
Finally in section \ref{sec:discussion} results are 
discussed with reference to the potentiality and limits of the
applied methodology.

\section{Data sources and methods}\label{sec:datamethods}
\subsection{Data sources}\label{subsec:data}
\begin{figure*}
\begin{center}
  \includegraphics[width=0.75\textwidth]{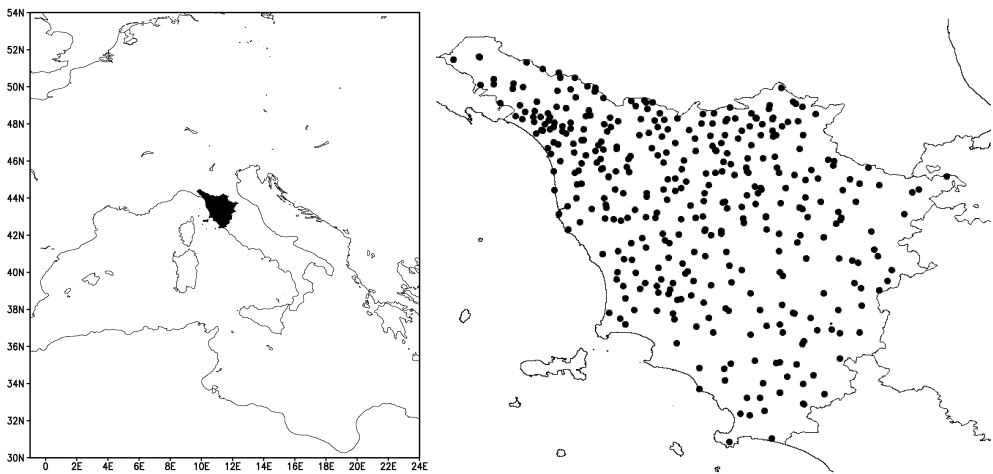}
\caption{Location of rain gauges (right side of the picture)-
         belonging to the Centro Funzionale Regionale 
         network in Tuscany (Central Italy, on left side of the picture).}
\label{fig:1}
\end{center}
\end{figure*}

The location of the rain-gauges belonging to
the Centro Funzionale Regionale (CFR) in Tuscany (Central Italy)
is shown in Figure \ref{fig:1}. 
Its establishment has been a 
long-term decision process involving several local 
institutions over more than 20 years. 
The network comprises 377 stations and encloses several basins 
over an area of about 23000 km$^2$ (yielding a density of about
one station every 60 km$^2$). More than 90\% 
of the stations are located below 810 meters. 
The biggest inter-station distance (in other words
the size or diameter of the point-set) is about 250 km.
These data make the geography of our network similar to
that studied by
\citeauthor{mazzarella2000} (\citeyear{mazzarella2000}).
\par Satellite imagery acquired by the Meteosat Second Generation (MSG-2) 
satellite in the infrared (IR) channel centered at 10.8 $\mu$m was used 
in this study as a proxy to detect and monitor cold clouds systems. 
The study period is July 2010 while the spatial and temporal data resolutions 
are 4.5 $\times$ 4.5 km$^2$ and 15 minutes, respectively.
A brightness temperature ($T_B$) threshold was used to identify cold cloud systems 
that are most likely to be associated with convective activity. \citeauthor{kolios2007} 
(\citeyear{kolios2007}, \citeyear{kolios2010}) used a $T_B$ of 228 K to best identify 
convective systems in the Mediterranean area based on a set of lighting data. 
The same temperature threshold of 228 K was used by \cite{morel2002} 
for assessing the climatology of the European MCSs, this value being very close to 
that of 221 K used by \cite{garcia2005} for Spain. This low-temperature threshold 
allows to investigate mostly anvil regions and  embedded areas of active deep 
convection (Johnson et al. 1990). In this study, a 228 K $T_B$ threshold value was 
chosen for identifying very deep convective events over Tuscany. 
Furthermore, discrimination between precipitating and non-precipitating cold cloud systems, 
previously subjected to $T_B$ threshold test, was performed using RADAR data provided by 
the DPCN (National Civil Protection Department) radar network. The data consist of a 
mosaic of instantaneous surface rainfall intensities (SRI) with a spatial and temporal 
resolution of 1 km and 15 minutes, respectively. Several daily precipitation amounts were tested; 
thresholds values of 1 mm, 2 mm, 5 mm, 10 mm, 15 mm, 20 mm, 25 mm, 30 mm, 35 mm per day
were used to analyze the different phenomenology linked to 
precipitation, from weak to moderate regimes.
\subsection{Methodology}\label{subsec:methodology}
\begin{figure}
\begin{center}
\includegraphics[angle=0,scale=0.35]{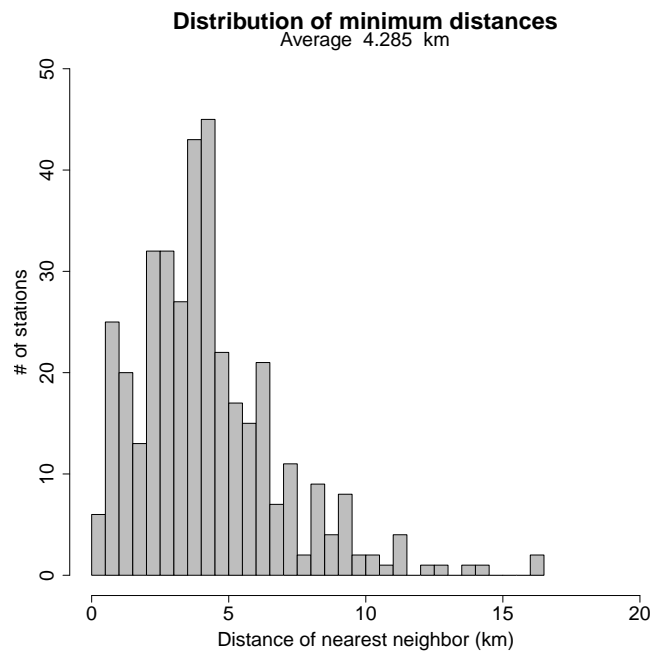}
\caption{Distribution of nearest neighbor's distances for each station in the point-set. 
         Average value is 4.2 km which is taken as $R_{min}$, the lower limit for the 
         regression of $\log(C(R))$ on $\log(R)$.}
\label{fig:2}
\end{center}
\end{figure}

While euclidean geometry deals with ideal geometric 
forms and assigns dimension 0 to points, 1 to lines 
and so on, fractal geometry deals with non-integer dimensions.
The fractal, or Hausdorff, dimension $D_H$ has been the most 
common used measure of the \textit{strangeness} of attractors 
of dissipative dynamical systems that exhibit chaotic 
behavior (\citeauthor{grassberger1983b} \citeyear{grassberger1983b}).
Since for experimental data the value of $D_H$ is difficult 
to determine using the box-counting algorithm (\citeauthor{strogatz1994} 
\citeyear{strogatz1994}), we computed the fractal dimension $D_2$
 of point-set using the  method proposed in 
\citeauthor{grassberger1983a} (\citeyear{grassberger1983a}; \citeyear{grassberger1983b}), 
as also found in the literature 
(\cite{korvin1990}, \cite{lovejoy1986}, \citeauthor{mazzarella2000} (\citeyear{mazzarella2000}),
\cite{olsson1996}).
In the case under examination we choose to use $D_2$ as a good approximation of $D_H$, since 
as stated in \cite{grassberger1983b},
$D_2 \le D_H$
and inequalities are rather tight in most cases.
\newline In the present study we compute the correlation dimension in a 2-dimensional 
space but in general, to obtain $D_2$ given a point-set $\{\mathbf{X}_i\}_{i=1}^{N}$ 
with $\mathbf{X}_i\in\mathbf{R}^n$, we have to
consider the correlation integral $C(R)$ that counts the number of pairs $\left\{\mathbf{X}_i,\mathbf{X}_j\right\}$
such that $\parallel\mathbf{X}_i-\mathbf{X}_j\parallel$ is smaller than a given threshold $R>0$, 
with $\parallel\cdot\parallel$ being the standard euclidean distance in $\mathbf{R}^n$.
In formulas:
\begin{equation}\label{eq:corrint}
C(R) = \frac{2}{N(N-1)}\sum_{i=1}^N \sum^N_{\substack{j=1 \\
j \ne i}} \Theta(R - \parallel\mathbf{X}_i- \mathbf{X}_j\parallel),
\end{equation}
where $\Theta$ is the Heaviside function and where $\frac{2}{N(N-1)}$ 
is the normalization factor so that $C(R)$ tends to $1$ for $R$ tending to infinite.
\newline If the rain-gauge network is a fractal then 
$C(R)$ grows like a power:
\begin{equation}
C(R) \varpropto R^{D_2},
\end{equation}
that is
\begin{equation}\label{eq:loglog}
\log(C(R)) \varpropto D_2\log(R).
\end{equation}
Therefore, one can derive $D_2$ from the regression coefficient of relationship (\ref{eq:loglog}).
\newline In order to determine the correlation dimension $D_2$ of the
rain-gauge network described previously, 
we computed the correlation function defined in equation 
(\ref{eq:corrint}), as described in \cite{lovejoy1986}, 
i.e. we determined the cumulative frequency distribution of the
inter-station distances for the total number of 377 stations.
The distances were determined by spherical trigonometry,
using geographic coordinates and ignoring elevations owing
to the smallness of the elevation with respect to the two horizontal
dimensions.
\newline For what concerns the values of the parameter $R$, as done 
in \citeauthor{mazzarella2000} (\citeyear{mazzarella2000}), we started 
in computing the inter-station 
distances defined in equation (\ref{eq:corrint}) from 1 km.
This value was gradually increased by a factor of 1.1 up to 250 km since, 
as expected by definition of $C(R)$ given in equation (\ref{eq:corrint}), for all 
$R\ge\text{size (area of interest)}$
the correlation integral $C(R)$ saturates to 1 and $\log(C(R))$ saturates to 0. 
\newline For experimental data the linear behavior of $\log(C(R))$ on $\log(R)$
is limited to a scaling region $S_R$, i.e. only for $R$ belonging to 
the interval $S_R=[R_{min},R_{max}]$
 (\citeauthor{strogatz1994} \citeyear{strogatz1994}). 
This happens because $C(R)$ is underestimated
from those points near the edge of the set so that the criteria to determine 
the bounds of $S_R$ need to be analyzed in each singular 
case (\citeauthor{liebovitch1989} \citeyear{liebovitch1989}).
According to the literature
(\citeauthor{forrest1979} \citeyear{forrest1979}; 
\citeauthor{grassberger1983a} \citeyear{grassberger1983a}; 
\citeauthor{korvin1990} \citeyear{korvin1990}), the upper limit $R_{max}$ is chosen equal to
 one third of the diameter of the area (about 80 km). 
In order to choose the lower limit $R_{min}$, we didn't perform any statistical
significance computation, 
since in our case the correlation 
coefficients are statistically significant at 99\% confident level for all $R\ge1$ km. 
Rather,
for each station we computed the distance of the nearest neighbor 
and took the average of this distribution
as the meaningful index of the points separation.
\newline In Figure \ref{fig:2} we plot this distribution; 
the average of nearest neighbor's distances is 
about 4.2 km and this value is considered as the lower limit $R_{min}$ 
meaningful for the regression.
\section{Results}\label{sec:results}
\begin{figure}
\begin{center}
  \includegraphics[angle=270,scale=0.35]{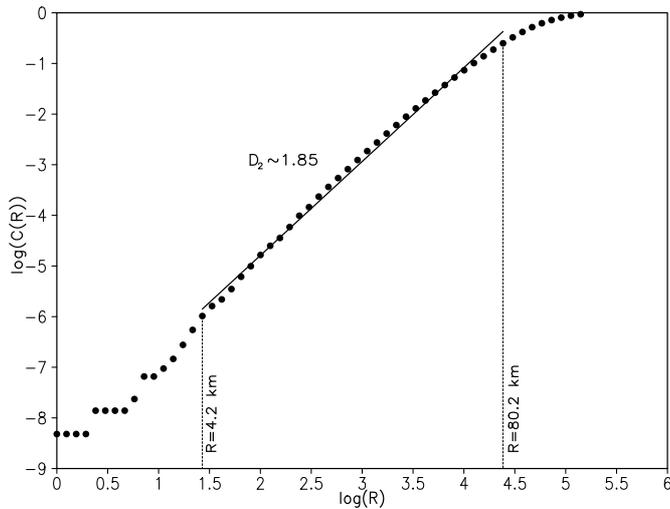}
\caption{Log-log plot of correlation integral $C(R)$ on $R$ with 
         scaling region delimited by $R_{min}=4.2$ km 
         and $R_{max}=80.2$ km. The corresponding slope of the 
         regression line which determines the 
         correlation dimension $D_2$ is equal to 1.85.}
\label{fig:3}
\end{center}
\end{figure}

The linear fitting between $\log(C(R))$ and $\log(R)$ within the scaling region $S_R$ bounded 
by $R_{min}=4.2$ km and $R_{max}=80.2$ km yields a slope, and thus a correlation dimension value
$D_2$ of 1.85. Figure \ref{fig:3} shows the results of this regression. 
The dimensional deficit $\delta$ of the network, defined
as the difference between the dimension of the embedded space
and $D_2$, is $(2-D_2)=0.15$.
\par
\begin{figure}
\begin{center}
  \includegraphics[angle=0,scale=0.35]{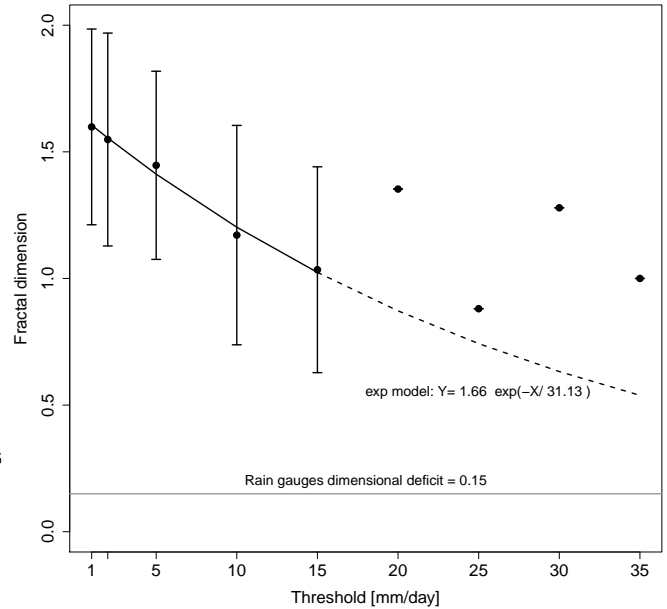}
\caption{Correlation dimensions of rainfall events (average values and standard deviations) 
         registered in Tuscany 
         in the month of July 2010 for different thresholds. 
         Horizontal gray line is $\delta$, the dimensional deficit of 
         rain-gauge network.}
\label{fig:4}
\end{center}
\end{figure}

 
The value of $\delta$ should be related to the dimension $d$ of rainfall phenomena.
From a climatological
point of view, the area of interest 
(central Italy) is mainly affected by convective storms or frontal systems, depending on the seasonality. 
Convective storms are of uppermost interest for our analysis since they are smaller, 
more or less separate rainfall areas displaying a considerable spatial variability and thus 
suitable for fractal analysis. They are typical of the warm season (from June to September roughly). 
The frontal storms are characterized by continuous rainfall areas of large spatial 
extensions and are typical in autumn and winter seasons.
\newline Using remote sensing and ground instruments described in section \ref{sec:datamethods} 
we collect data for every day in July 2010. 
First step is to select all the pixel in the MSG-2 15-minutes dataset having a brightness 
temperature below 228 K so that we can obtain a point-set (i.e. pixel-set) of 
potential precipitation cells (\citeauthor{kolios2010} \citeyear{kolios2010}; 
\citeauthor{morel2002} \citeyear{morel2002}). 
Secondly, to assign a rain amount to the selected pixels we use the radar data. 
For each selected pixel in the MSG 15-minutes dataset we retrieve the surface rainfall intensities (SRI)
as estimated by the RADAR data provided by the DPCN (National Civil Protection Department).
Finally for each day of July 2010 we add all the 96 daily images (for each day we 
have one image every 15 minutes) and obtain a daily estimate of precipitation amount.
Rain estimates were processed in order to compute the correlation dimension 
using the method detailed in section \ref{sec:datamethods}.
In figure \ref{fig:4} we plot the correlation dimensions of rainfall events registered 
in the month of July 2010 in Tuscany. Daily rainfall events were divided on the basis of 
prescribed thresholds, chosen equal to 
1 mm, 2 mm, 5 mm, 10 mm, 15 mm, 20 mm, 30 mm, and 35 mm.
For each threshold, Figure \ref{fig:4} shows the average and standard deviation values 
of correlation dimension of the rainy pixel-set for those days having a significant number 
of points that registered an amount of precipitation above the threshold.
%
\section{Discussion}\label{sec:discussion}
The present study achieved the issue to estimate
the areal sparseness of the monitoring rain-gauge
network belonging to the CFR
owned by Tuscany Administration by means of the fractal (correlation)
dimension $D_2$.
In Table \ref{tab:1}, we compare this value with dimension $D_2$ found in other,
similar studies in the literature.
Except the cases of Australia and Canada, where the dominance of inhabited areas along the coast
lowers the value of the fractal dimension, our $D_2$ value is in good agreement with the others.
\begin{table}
\begin{tabular}{llll}
\hline\noalign{\smallskip}
Reference & \# of points & Area coverage & $D_2$  \\
\noalign{\smallskip}\hline\noalign{\smallskip}
$-$ \cite{lovejoy1986} & 9563 & global land & 1.75 \\
 & 3593 & France & $\simeq$1.8 \\
 & 414 & Canada & $\simeq$1.5 \\
$-$ \cite{korvin1990} & $\simeq$65000 & Australia & 1.42 \\
$-$ \cite{tessier1994} & 7983 & global land & 1.79\\
$-$ Olsson & 230 & $\simeq$ 10000 km$^2$ & $\simeq$2 \\
and Niemczynowicz (1996) & & & \\
$-$ Mazzarella & 215 & $\simeq$ 38000 km$^2$ & 1.84 \\
and Tranfaglia (2000) &  &  &  \\
 & 300 & $\simeq$ 38000 km$^2$ & 1.89 \\
$-$ Present study & 377 & $\simeq$ 23000 km$^2$ & 1.85 \\
\noalign{\smallskip}\hline
\end{tabular}
\caption{Comparison of $D_2$ computed in the present study with
         other correlation dimensions found in literature. 
         It is also reported the
         bibliography references, the number of points taken into account and
         the extent of geographical area.
         }\label{tab:1}
\end{table}
However, we have to point out that the computed correlation dimension $D_2$
must be handled with care because, according to the
Tsonis criterion (\citeauthor{tsonis1994} \citeyear{tsonis1994}), 
the minimum number $N_{min}$ of points 
required to produce a correlation integral with no more than an
error $Err$ (normally $Err = 0.05D_E)$ is approximately
\begin{displaymath}
N_{min}\varpropto 10^{2+0.4D_2}
\end{displaymath}
which, in our case, means $N_{min}\simeq 600$ whereas we have 377 stations.
\newline\indent In order to analyze the dimensional deficit $\delta=0.15$, we need to compare it
with the fractal dimension of rainfall events, as done in the previous section.
As for the precipitations fallen in July 2010, even the more intense ones were
characterized by a fractal dimension $d>0.6$, much greater than the dimensional
deficit $\delta=0.15$. We stress that the used rain data (see Figure \ref{fig:4}) are
\textit{independent} from the rain-gauge network used to determine the fractal
dimension $D_H=1.85$. Therefore, these preliminary results allow us to state with a good 
confidence that our rain-gauge network was precise enough to record all precipitation events
occurred in July 2010. Empirically we can suppose that the fractal dimension goes to zero as
the threshold increases; this because intense precipitation events (at least thermo-convective ones)
are more scattered. On the other hand light rains are more homogeneus and then associated to a (decreasing) linear
trend for small thresholds. We then choose to fit the values plotted in
Figure \ref{fig:4} with an exponential model which is almost linear for small values of  
$x$ and decreases to zero as $x$ tends to infinity.
We can then suppose that
\begin{displaymath}
y=Ae^{-x/x_0}
\end{displaymath}
where the independent variable $x$ on the $x$-axis represents the daily amount of rain, 
$y$ represents the fractal dimensions
and the parameters are $A=1.657$ and $x_0=31.133$. 
The model is calibrated using the just the first values of $x$ 
(thresholds from 1 mm/day up to 15 mm/day, continuos line in Figure \ref{fig:4}), since above we don't have any 
significant statistics (just two days registering at least 20 pixels with a precipitation 
above 20 mm and one day registering at least 20 pixels with a precipitation above 25 mm).
\newline The exponential regression intersects $\delta=0.15$ for 
\begin{displaymath}
x_\delta=-x_0\ln\left(\frac{\delta}{A}\right)
\end{displaymath}
that is $x_{0.15}\simeq 75$ mm/day.
In other words our data suggests that rainfalls with daily amount equal or above $75$ mm/day
might correspond to a fractal dimension $d<\delta$, so that these events could not be detected. 
This value is based, by construction, on the remote sensed data and ground instruments 
and on the phenomenology of rainfall events. Ongoing efforts are directed toward the improvements 
of the accuracy of instruments and toward the calibration of the algorithms.
\newline\indent Further studies are required to investigate the relationship between correlation dimension
$D_2$ of the observing network (but we are rather interested in the dimensional deficit $\delta$) 
and the dimensions $d$ of rainfall events. Firstly the most important improvement of the research is to expand 
the statistics of precipitating events considering several months for, at least, a couple of years. 
Moreover it would be interesting to take into consideration the fall/winter precipitations, which are mainly 
associated with cold and warm fronts, to evaluate the different behavior of dimension $d$ and
check if, for some thresholds, it drops below $\delta$.
%

%
\begin{acknowledgements}
MSG imagery is copyright of EUMETSAT and was made available by the EUMETSAT on-line Archive.
We thank National Department of Civil Protection (DPCN) for providing weather radar data.
The authors are grateful to Andrea Antonini and Stefano Romanelli for preprocessing
and providing remote sensed data.
In addition special thanks go to Melissa Morris for the revision of the text.
Partial financial support by Regione Toscana is gratefully acknowledged.
\end{acknowledgements}

\bibliographystyle{spbasic}     
\bibliography{./bibliocapecchi}   

\end{document}